\documentclass[prb,twocolumn,showpacs,preprintnumbers,amsmath,amssymb,longbibliography]{revtex4-1}
\usepackage{graphicx}
\usepackage{dcolumn}
\usepackage{bm}
\usepackage{times}
\newcommand{\BiTe}{Bi$_{2}$Te$_{3}$ }
\newcommand{\BiTeSe}{Bi$_{2}$Te$_{2}$Se }
\newcommand {\be}{\begin{equation}}
\newcommand {\ee}{\end{equation}}
\newcommand {\bea}{\begin{eqnarray}}
\newcommand {\eea}{\end{eqnarray}}
\begin{document}

\title{Connecting thermoelectric performance and topological-insulator behavior: \BiTe and \BiTeSe from first principles}

\author{Hongliang Shi, David Parker, Mao-Hua Du and David J. Singh}

\affiliation{Materials Science and Technology Division,
Oak Ridge National Laboratory, Oak Ridge, Tennessee 37831-6056}

\date{\today}

\begin{abstract}
Thermoelectric performance is of interest for numerous applications such as waste heat recovery and solid state energy conversion, and will be seen to be closely connected to topological insulator behavior.
In this context we here report first principles transport and defect calculations for \BiTeSe in relation to Bi$_{2}$Te$_{3}$.  The two compounds are found to 
contain remarkably different electronic structures in spite of being isostructural and isoelectronic.  We discuss
these results in terms of the topological insulator characteristics of these compounds.
\end{abstract}

\maketitle

\section{Introduction} 

Thermoelectric performance is typically quantified in term of a dimensionless figure-of-merit $ZT$, 
given by the following expression:
\bea
ZT &=& \frac{S^2 \sigma T}{\kappa}
\eea
Here $S$ is the Seebeck coefficient or thermopower, $\sigma$ is the electrical conductivity, $T$ the absolute temperature, and $\kappa$
the thermal conductivity.  The expression shows that for good performance one desires both high electrical conductivity and Seebeck coefficient,
but these are difficult to obtain simultaneously due to opposite dependencies on carrier concentration.  Hence thermoelectric performance
is a counter-indicated property of materials that does not commonly occur, and determining and optimizing a usable high performance thermoelectric material remains a difficult challenge.

Thermoelectric performance is of considerable engineering and technological importance due to the many
potential applications of this technology, which include vehicular exhaust waste heat recovery, energy harvesting, heating and cooling, and
solid state energy conversion.  In all of these applications higher thermoelectric performance would be extremely beneficial for enhanced device performance.  Currently,
there are relatively few thermoelectrics with ZT values above unity, the minimum necessary for a thermoelectric to be considered high performance.  This has greatly limited the
utility of thermoelectrics, leading to substantial efforts aimed at raising ZT.

Presently the thermoelectric most employed in applications is Bi$_{2}$Te$_{3}$, a narrow gap semiconductor which shows optimized
$ZT$ figures of approximately unity at ambient temperature.  It is presently used primarily in niche applications.

Of great consequence for potential applications, at temperatures above 300 K, the performance of \BiTe degrades rapidly due to bipolar conduction, or the excitation of 
\begin{figure}[b!]
\vspace{-1.3cm}
\includegraphics[width=0.8\columnwidth,angle=270]{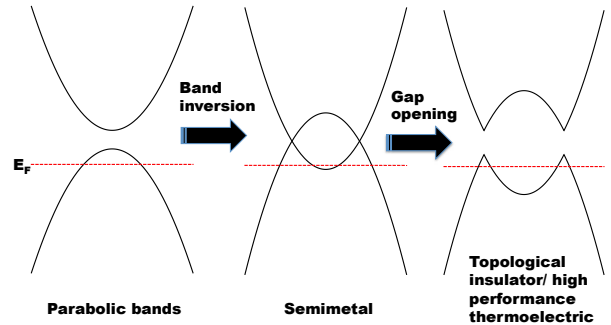}
\vspace{-1cm}
\caption{Depiction of the effects of spin-orbit coupling in generating topologically insulating, potential high performance thermoelectrics, by means
of opening of a gap in the electronic structure, with associated non-parabolicity. A material doped $p$-type is depicted.}
\end{figure}carriers of both positive and negative charge.  This causes the thermopower to decrease with increasing temperature, the opposite
of the usual situation, and in addition causes large increases in the electronic thermal conductivity.   Both of these effects are destructive
for thermoelectric performance, as suggested by Eq. 1.  These effect generally occurs when the semiconductor band gap (about 0.15 eV in Bi$_{2}$Te$_{3}$) is not sufficiently large relative
to the device operating temperature.  In the absence
of bipolar conduction, $ZT$ is a strongly increasing function of increasing temperature, with performance ultimately limited only by the decomposition or melting point of the material.

Bi$_{2}$Te$_{3}$,  therefore, could be an extremely high performance thermoelectric at temperatures of 400 to 500 K,  if only its band gap were somewhat larger.    This would be of great practical importance given that two major potential applications - exhaust waste heat recovery and solid state thermophotovoltaic conversion - operate at temperatures around 500 K.  Part of this work will explore a potential scenario for achieving this.

While \BiTe has been known as a high performance thermoelectric for several decades,  it also 
forms the basis for a family of topological insulators (TI) (Bi,Sb)$_{2}$(Te,Se)$_{3}$  \cite{jia}.  Many have observed a connection
between these two properties, and various explanations proposed; perhaps the simplest one
is the observation that good thermoelectrics are usually heavy atomic mass, small band gap semiconductors, as
the heavy atom helps to induce low lattice thermal conductivity, as well as the TI band inversion (via spin-orbit coupling),  
and the small band gap high carrier mobility.  A floor on the degree of band inversion necessary to produce TI is set by the band gap, presumably making large band gap TI's less common.

However, not every, or even a significant fraction of heavy mass small gap materials, are good thermoelectrics
or good TI materials.  Furthermore, some materials without heavy mass atoms or small gaps are excellent thermoelectrics,
such as Mg$_{2}$(Si,Sn) and Si-Ge.   In addition, from an electronic point of view TI behavior is of interest for an undoped
material (where the Fermi energy is in the gap) while high thermoelectric performance is usually observed with the Fermi energy
doped into the bulk bands.

Like thermoelectric performance, topologically insulating behavior is of considerable practical importance due to its potential for technological 
applications, such as memory applications for computers \cite{mellnik}.  Here we show a clearer connection between topologically insulating behavior and thermoelectric performance.  Briefly, we will see that complex, non-parabolic band structures are 
favorable both for TI behavior and high thermoelectric performance. 
In this work we will
see that two materials studied as topological insulators - \BiTe and \BiTeSe - appear to have very complex band structures that are
in general highly beneficial for thermoelectric performance.
These complex band structures are related to TI behavior, as the band inversion necessary for this generally creates complex band structures 
not typically describable in terms of the usual anisotropic effective mass approximation.  Remarkably, the two compounds are very different in the 
near band edge electronic structures leading to very different transport behavior.

In Figure 1 we depict schematically the effects of spin-orbit coupling in producing the complicated band structures just mentioned. Briefly, the band inversion central to
TI behavior is induced by spin-orbit coupling, which then opens a gap at the points where the bands would otherwise cross. As depicted in Figure 1, this generally
leads to non-parabolic behavior, often with near-linear Kane band-type dispersions.  Thus a single, parabolic, non-degenerate band edge, as shown in the
left side of this figure, evolves into a non-parabolic, complex, degenerate band edge, as is often observed in high-performance thermoelectrics.  More detailed 
discussions of these effects can be found in Refs. \onlinecite{hsieh,chen_top,zhang2}.

Bi$_2$Te$_3$ exhibits the band inversion required for topologically insulating behavior, but is inconvenient for studying TI.
This is because of its small band gap and small defect formation energies, which mean that low bulk electrical conductivity - a prerequisite for observing the topologically protected surface states - is difficult to attain.  This is due both to large bipolar conduction, in the lightly doped intrinsic regime and large band conduction (in the heavily doped extrinsic regime favored by the low vacancy formation energies).  This small band gap also presents a substantial hindrance to thermoelectric applications above room temperature, as bipolar conduction is highly destructive to thermoelectric performance.

Perhaps with this small band gap in mind, significant recent efforts  have been focused on the topologically insulating properties of the isoelectronic and isostructural Bi$_2$Te$_2$Se, (experimental band gap of  $\sim$ 0.30 eV) where one of the Te layers (see Fig. 2) is entirely replaced with Se.  Relatively recently, \cite{jia} low bulk conductivity single crystals of this material were synthesized and studied, a major step forward towards the experimental verification of the surface states.  To date, however, relatively little attention has been directed to the {\it thermoelectric} properties of this compound. Indeed, its larger band gap suggests a propensity for thermoelectric performance at temperatures above those of Bi$_2$Te$_3$.  Disordered alloys near this composition appear to show some potential for thermoelectric performance at higher temperatures, but not as high as if the low T behavior of \BiTe could be extended to higher T.

\begin{figure}
\includegraphics[width=0.8\columnwidth,angle=0]{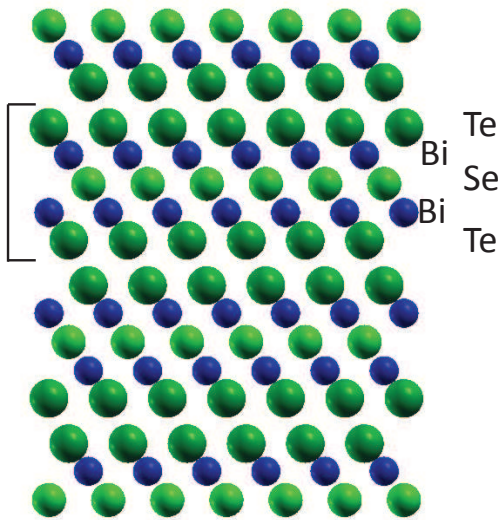}
\caption{Depiction of the crystal structure of Bi$_2$Te$_2$Se,
showing the layer stacking along the rhombohedral $c$-axis. The
bracket ``[" indicates a single Bi$_2$Te$_2$Se layer with Se in the
central plane. The atomic positions are taken from the relaxed structure.}
\label{struct}
\end{figure}

Bi$_2$Te$_2$Se forms with a structure closely related to that of Bi$_2$Te$_3$.
In particular, as shown in Fig. \ref{struct}, it has a tetradymite type
rhombohedral (spacegroup 166)
crystal structure, consisting of Bi$_2$Te$_2$Se layers stacked
along the $c$-axis and separated by van der Waals gaps. These
Bi$_2$Te$_2$Se layers are the same as the Bi$_2$Te$_3$ layers comprising
Bi$_2$Te$_3$ except that the central Te plane is replaced by a Se plane.
\cite{wiese,nakajima,misra}
Presumably this particular substitution is favored by the fact that placing Se on
this site places this more electronegative atom on the site with the
best metal coordination.

The growth of high quality crystals of this material has recently been
perfected,  enabling experimental study of its topological insulating
behavior. \cite{jia,ren}
The compound naturally forms $n$-type from the melt.
However, recent experiments
have shown control of the carrier concentration using Sn doping (which
introduces mid-gap states) and excess Bi.
\cite{jia,ren,jia2}

The thermoelectric properties of Bi$_2$Te$_2$Se were recently investigated
by Fuccillo and co-workers. \cite{fuccillo}
There has also been recent theoretical and experimental work on the
potential performance of nanostructured
Bi$_2$Se$_3$ and its alloys with Bi$_2$Te$_3$.
\cite{parker-bi2se3,liu}
These studies find that Bi$_2$Se$_3$ and compounds between
it and Bi$_2$Te$_3$ can have higher $p$-type thermopowers than Bi$_2$Te$_3$
especially at temperatures above the operating temperature of Bi$_2$Te$_3$, suggesting a 
propensity for enhanced $p$-type performance at these temperatures.  These studies also
suggest that reasonable thermoelectric performance is possible with
reduction of the thermal conductivity by nanostructuring.

\section{Electronic structure calculations} 

We performed the present calculations using Boltzmann transport theory
with the first principles electronic structure, employing the
constant scattering time approximation (see Ref. \onlinecite{parker-PbSe} for a detailed
description of this approximation.) 
The BoltzTraP code \cite{boltztrap} was used for these transport calculations, and
the electronic structure obtained using the modified Becke-Johnson
potential of Tran and Blaha (TB-mBJ).
\cite{tran}
This potential gives very much improved band gaps for simple semiconductors
and insulators as compared to standard density functionals.
\cite{tran,koller,singh-1,kim-mbj,singh-2,camargo}
These calculations employed the general potential linearized
augmented planewave (LAPW) method, \cite{singh-book}
as implemented in the WIEN2k code. \cite{wien2k}
Experimental lattice parameters,
($a$=4.3792 \AA, $c$=30.481 \AA{} for Bi$_2$Te$_3$ and
$a$=4.305 \AA, $c$=30.00 \AA{} for Bi$_2$Te$_2$Se) \cite{misra}
were used. The free internal atomic coordinates were determined
by total energy minimization using the local density approximation (LDA).

The LDA was used because it
was found to yield better structural and vibrational
properties for Bi$_2$Te$_3$ than
generalized gradient approximations when used with fixed lattice parameters
for Bi$_2$Te$_3$. \cite{chen-bi2te3}
The structure relaxation was done treating relativity at the scalar
relativistic level, as relaxation including spin-orbit coupling is not supported in WIEN; the effect of this omission is likely minimal.
All the other reported results include spin-orbit coupling,
including the electronic structures and transport properties.
Well converged basis sets, defined by a
cut-off $RK_{max}$=9.0 for the planewave vector plus local
orbitals for the semi-core $d$ states were used. Here $k_{max}$
is the planewave cut-off and $R$ is the sphere radius, which was
taken as 2.5 bohr for all atoms.

The calculated band gaps are 0.14 eV for Bi$_2$Te$_3$
and 0.22 eV for Bi$_2$Te$_2$Se. Thus the band gap of Bi$_2$Te$_2$Se
is significantly larger than that of Bi$_2$Te$_3$, although
still smaller than that of the  higher temperature thermoelectric
PbTe (0.36 eV, by a similar method). 
\cite{ekuma}
Experiment also shows a similar increase in band gap when Se is added to 
Bi$_2$Te$_3$, i.e. the optical absorption edge is reported to 
increase from
$\sim$ 0.15 eV in Bi$_2$Te$_3$ to $\sim$ 0.30 eV at a composition
Bi$_2$Te$_2$Se. \cite{greenaway}

Hinsche and co-workers \cite{Hinsche} reported Boltzmann transport
calculations for Bi$_2$Te$_3$. They found results similar to ours
for the thermopower and conductivity, and in particular found
better conductivity for the in-plane directions and higher values
of the thermopower for $p$-type doping.

We present the calculated band structure for both materials in Figure 3.
\begin{figure}
\includegraphics[width=0.8\columnwidth,angle=0]{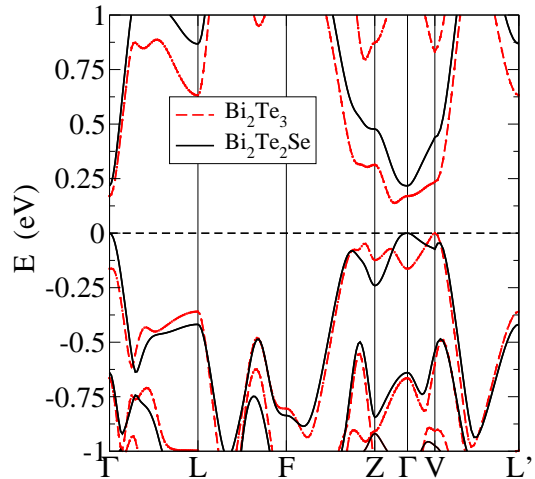}
\caption{The calculated band structure of Bi$_2$Te$_3$ and Bi$_2$Te$_2$Se.  We have set the energy 0 to be the valence band maximum for both materials.  The point ``V" refers
to the approximate location of the valence band maximum of Bi$_2$Te$_3$ - (2/5,2/5,1/3) in the rhombohedral basis, and L' to the point (0,0,-1/2) in the same basis.}
\end{figure}
Although some of the features, such as the valence bands more than 0.5 eV below the valence band maximum, are similar the fine details of the electronic structure are in fact very different.  For example, both band extrema for Bi$_{2}$Te$_{3}$ are at off-symmetry locations (the Bi$_{2}$Te$_{3}$ valence band maximum V is approximately (2/5,2/5,1/3) in the rhombohedral basis, a non-symmetry point), while both extrema for \BiTeSe are at the $\Gamma$ point.  This has important implications for thermoelectric performance as the increased band degeneracy of Bi$_{2}$Te$_{3}$ is one likely contributor to its high thermoelectric performance.  The valence band of Bi$_{2}$Te$_{3}$ has two subsidiary maxima located near the Z point, while \BiTeSe has two subsidiary valence band local maxima located at different points.

One plausible question to ask, given the argument of the Introduction for the correspondence between the complex band structures favorable for both thermoelectric performance and topological insulators, is the 
relationship of the above band structures to spin-orbit coupling. In order to address this question we present in Figure 4 the results of calculations in which the effective strength of the spin-orbit coupling is varied from
zero to unity (the fully spin-orbit case).  
\begin{figure}
\includegraphics[width=0.8\columnwidth,angle=90]{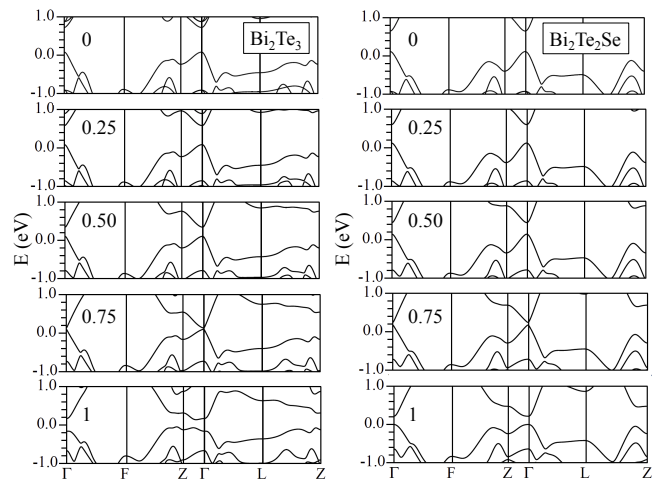}
\caption{The calculated band structures of Bi$_2$Te$_3$ and Bi$_2$Te$_2$Se with the spin-orbit coupling included
in strengths (relative to the actual physical value) of 0, 0.25, 0.5, 0.75 and unity.}
\end{figure}
As the plots indicate, without spin-orbit both materials are direct gap semiconductors with band edges at the $\Gamma$ point
and comparatively parabolic bands.  In both cases, however, as the spin-orbit interaction is turned on the band gap decreases
radically until in the strength=0.75 plot the gap is very small - less than a tenth of an eV, and the bands become visibly non-parabolic. Finally, 
as in the right-hand panel of Figure 1, when the full strength of spin-orbit is applied a new gap opens up - between the Z and $\Gamma$
points for \BiTe but returning to the $\Gamma$ point for Bi$_2$Te$_2$Se, and these band structures do appear
to be comparatively non-parabolic.  Note also that the motion of the band edges in \BiTe away from the $\Gamma$ point with the
advent of spin-orbit automatically implies a more complex Fermi surface structure due to the associated degeneracy, irrespective of the parabolicity
of the bands.

By way of comparison, the band structure of \BiTeSe is in fact much more similar to that of Bi$_{2}$Se$_{3}$, which also has both extrema at the $\Gamma$ point, than that of Bi$_2$Te$_{3}$, despite being closer to the latter compound compositionally.  Further insight can be obtained by plotting the isoenergy surfaces of both materials, as presented in Figure 5.  
\begin{figure}
\includegraphics[width=1.0\columnwidth,angle=0]{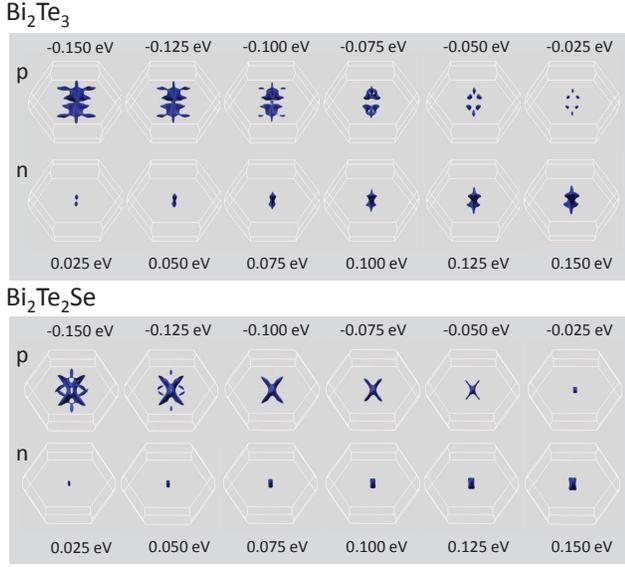}
\caption{The calculated isoenergy surfaces of Bi$_2$Te$_3$ and Bi$_2$Te$_2$Se.  The energies given represent the isoenergy value, relative to the respective band extrema.}
\end{figure}
For both materials, for $p$-type doping a highly anisotropic, non-parabolic behavior is evident.   Recall that in a parabolic approximation the isoenergy surfaces takes the form of ellipsoids of
revolution, even if effective mass anisotropy is considered.  Neither of these materials exhibits a $p$-type Fermi surface at all resembling an ellipsoid; for Bi$_{2}$Te$_{3}$ at the smallest energies a distinct triangular shape appears, followed at increasing binding energy by a bell-like structure and ultimately augmented with planar ``wings".  The shape is very different for Bi$_{2}$Te$_{2}$Se, with the initial VBM at $\Gamma$ rapidly evolving into an ``X" shaped figure (note that there are in fact 6 subsidiary extrema in this structure), which is then followed by a ring-like feature.  

All of these deviations from spherical, or ellipsoidal, shapes can be seen to be beneficial for thermoelectric performance.  For a given volume (in this case effectively carrier concentration), a sphere has the minimum surface area (in this case, effectively density-of-states [DOS] ), and therefore minimum thermopower, since in the degenerate limit the thermopower is proportional to the  DOS mass.  Hence all deviations from a spherical isoenergy surface enhance the thermopower, and the greater the deviation the greater the enhancement.   An example of this effect can be found in Ref. \onlinecite{chen}.  While a detailed quantitative comparison between the two materials on this basis is not readily available, we may state with some confidence that {\it both} materials, when doped $p$-type will benefit from the anisotropy of the electronic structure.

With regards to $n$-type, here the situation is substantially different.  While Bi$_{2}$Te$_{3}$ still affords a substantially anisotropic isoenergy surface, with a discus shape evolving out of a non-$\Gamma$ point extremum, for \BiTeSe there is only a single $\Gamma$-centered, relatively cylindrical extremum, and this cylindrical shape is notably ``closer" to a spherical shape than that of \BiTe.  Hence we expect, and will later see, diminished $n$-type performance for \BiTeSe relative to Bi$_{2}$Te$_{3}$. 

We note that all band structures are significantly different from the ``pudding-mold" band structure proposed by Kuroki et al \cite{kuroki} as an explanation for the simultaneous
occurrence of high thermopower and electrical conductivity in the cobalt ate Na$_x$CoO$_2$. In that band structure a flat upper portion provides the large density-of-states
necessary for a high Seebeck coefficient, while a dispersive portion connecting to this provides a light band which favors high conductivity.  Here \BiTe in particular, from Figs. 3 and 5, contains near-degenerate band edges
resulting from its complex isoenergy surfaces that allow it to attain high conductivity without sacrificing thermopower, a distinct scenario from that of Ref. \onlinecite{kuroki}.

Although it is not immediately apparent from the plots, the iso-energy surfaces reflect the rhombohedral symmetry, with the off-symmetry valence band maximum for \BiTe six-fold degenerate and the conduction band minimum
located on the trigonal axis two-fold degenerate. For $p$-type Bi$_{2}$Te$_{2}$Se, the ``X" emanating from the $\Gamma$ point (beginning at -0.05 eV) actually comprises six ``arms", as two of the ``arms" are hidden by the projection.  
 
\section{Boltzmann transport calculations}

Following the electronic structure calculations, we performed Boltzmann transport calculations of the doping and temperature-dependent thermopower
and electrical conductivity, within the ``constant scattering time approximation", which shows substantial success in describing thermopower of a large number of materials.   Within this theory of diffusive transport the expressions for the
thermopower and conductivity are
\begin{equation}
S={ {\int {\rm dE}\sigma(E)(E-E_F)f'(E-E_F)}
\over{\int {\rm dE}\sigma(E)f'(E-E_F)}} ,
\end{equation}

\noindent and

\begin{equation}
\sigma= \int {\rm dE}\sigma(E)f'(E-E_F) ,
\end{equation}

\noindent where $f'$ is the energy derivative of the Fermi function and
$\sigma(E)$ is the energy dependent transport function related to
conductivity, $N(E)<v^2(E)\tau(E)>$, $N(E)$ the density of states,
$v^2$ the square of the component of the band velocity on the direction
of the interest (i.e. $v_x$ for conductivity along direction $x$,
making $v^2$ a rank 2 tensor, like the conductivity), $\tau$
the inverse scattering rate. The constant scattering time approximation
is the neglect of the energy (but not doping
or temperature) depedence of $\tau$, so that the transport
function becomes $N(E)<v^2(E)>\tau$, where $<v^2>$ is the average
Fermi velocity (still a rank 2 tensor).

\begin{figure}
\vspace{-0.55cm}
\includegraphics[height=1.15\columnwidth,angle=270]{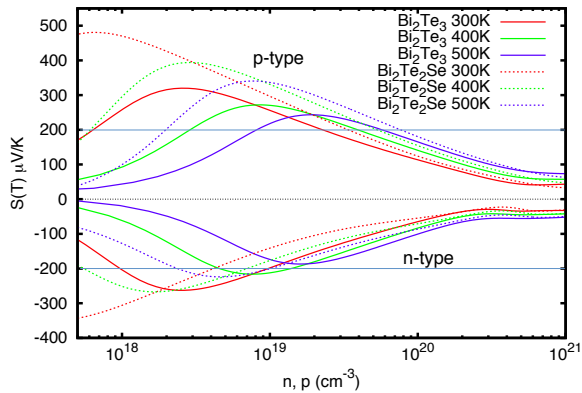}
\vspace{-1.2cm}
\caption{Conductivity averaged Seebeck coefficient as a function
of carrier concentration for Bi$_2$Te$_3$ (solid lines) and Bi$_2$Te$_2$Se
(dashed lines) for $p$-type (above zero line) and $n$-type doping
(below zero line) levels.  The horizontal blue lines indicate a thermopower magnitude of 200 $\mu$V/K, generally the minimum
necessary for a material to be a high performance thermoelectric.}
\label{seebeck}
\end{figure}

With these preliminaries completed, we move to the calculated quantities of interest.  In Figure 6 we present
the thermopower for the two materials, at three temperatures - 300, 400 and 500 K.  Note that due to the anisotropy of 
the electronic structure we have depicted the conductivity weighted thermopower, as would be
observed in the polycrystalline sample typically measured in the experiment.  For p-type, one notes
that the thermopower is significantly larger for \BiTeSe than for Bi$_{2}$Te$_{3}$, as a function of carrier concentration, for all
three temperatures.  This reflects the differing electronic structure of these two materials, as well as the larger calculated
band gap of Bi$_{2}$Te$_{2}$Se. At all three temperatures, $p$-type \BiTeSe displays a substantial range of carrier concentration where the thermopower 
is larger than 200 $\mu$V/K.  As we have noted elsewhere \cite{parker_Ag}, the Wiedemann-Franz relation essentially necessitates a thermopower magnitude of
200 $\mu$V/K or greater for a high performance thermoelectric; it is worth noting that this is the 300 K thermopower of optimally
doped Bi$_{2}$Te$_{3}$.  For n-type the thermopower of \BiTeSe appears inferior to that of Bi$_{2}$Te$_{3}$, even with the larger band gap, presumably
due to the less anisotropic, and hence less non-parabolic electronic structure.  We therefore focus on $p$-type behavior in the following.

The benefits of \BiTeSe relative to Bi$_{2}$Te$_{3}$ in the $p$-type thermopower should {\it not}, however, necessarily be taken as quantitative evidence for likely better, or even equal, thermoelectric performance
in Bi$_{2}$Te$_{2}$Se.    In order to assess this we plot the average electrical conductivity {\it versus} thermopower at 300 K in Figure 7.
Figure 7 reveals that in the $p$-type (right hand side of plot) region of thermopower around 200 $\mu$V/K, the two materials have virtually identical $\sigma/\tau$, which would indicate comparable thermoelectric transport, if the scattering times are equal.   The same behavior is evident at 500 K (not shown).  Note that in this comparison we are {\it not} referring to the bottom portion of the graphs, near where the thermopower transitions from positive to negative.  This region is firmly within the bipolar regime, well below optimal doping, and for which thermoelectric performance is generally poor. Instead we refer to the linear region adjacent to the legend, which is likely near where optimal performance would be found.

The isoenergy surfaces for $p$-type \BiTeSe appear to be somewhat less anisotropic than for Bi$_{2}$Te$_{3}$, which may explain why the thermopower benefits versus carrier concentration do not remain when compared to $\sigma/\tau$.  With regards to $\tau$, the scattering times may not be equal, given that in one sample of the line compound \BiTeSe  disorder \cite{jia} of order 5 percent was observed on the Te/Se sites, which would tend to decrease scattering times and hence electrical conductivity.   Optimal electrical conductivity in \BiTeSe therefore may necessitate extremely careful sample preparation in order to minimize this effect.

For a further comparison, in Fig. 8 we depict the calculated power factor $S^{2}\sigma/\tau$ (with respect to an average, unknown scattering time) at 300 K for both materials, for $p$-type and $n$-type, as a function of carrier concentration (carriers per unit cell).  The plot depicts comparable behavior for $p$-type, consistent with the behavior in Figures 6 and 7, noting that shorter scattering times in Bi${2}$Te$_{2}$Se may degrade the performance of this material relative to that of \BiTe. For $n$-type this figure suggests, consistent with the other figures, that Bi$_{2}$Te$_{2}$Se performance will significantly lag that of Bi$_{2}$Te$_{3}$.

\begin{figure}
\includegraphics[height=0.9\columnwidth,angle=270]{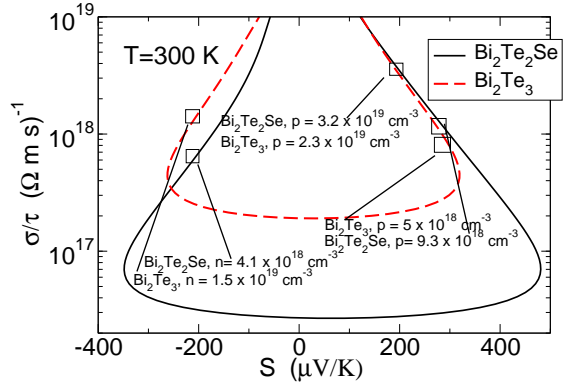}
\caption{Average 300 K conductivity versus Seebeck coefficient for Bi$_2$Te$_3$ (dashed lines) and Bi$_2$Te$_2$Se.}
\end{figure}
Returning to Figure 6, $p$-type Bi$_{2}$Te$_{3}$ shows doping levels where the thermopower is above 200 $\mu$V/K at temperatures above 300 K, where thermoelectric performance is usually believed to deteriorate.  This is most significant at 400 K but is true even at  500 K.   This means good thermoelectric performance may obtain at these temperatures.  Actual results, particularly at 500 K, will depend sensitively on the exact value of the band gap at these temperatures, as well as on any differences in hole and electron scattering times.  Performance would likely be optimized at dopings significantly heavier than those (about $ p = 2 \times 10^{19} $ cm$^{-3}$) used for commercial Bi$_{2}$Te$_{3}$.  This is necessary to minimize bipolar conduction.  At 400 K this doping level is  approximately 4 $\times 10^{19}$ cm$^{-3}$ and at 500 K it is 5.8 $\times 10^{19}$ cm$^{-3}$.  Due to the close proximity of the bipolar regime performance will rapidly degrade at dopings below these.  For $n$-type there is no such region of extended higher temperature performance for either \BiTe or Bi$_{2}$Te$_{2}$Se.
\begin{figure}
\includegraphics[height=0.9\columnwidth,angle=270]{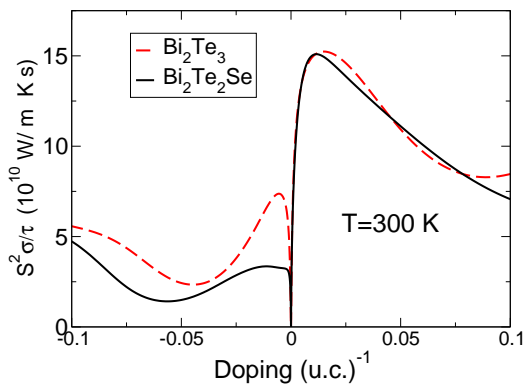}
\caption{Average 300 K power factor S$^{2}\sigma/\tau$ versus doping level for Bi$_2$Te$_3$ (solid lines) and Bi$_2$Te$_2$Se.}
\end{figure}

Figures 6, 7 and 8 together suggest that the likelihood of \BiTeSe performance exceeding that of Bi$_{2}$Te$_{3}$ is fairly low, even at the elevated temperatures where its larger band gap would be expected to be of advantage.  This has implications for the ongoing search for technologically useful thermoelectrics in the 400 to 500 K range, in particular suggesting that a larger band gap cannot necessarily be considered a panacea for  achieving high thermoelectric performance.  In this case it is the less favorable electronic structure of Bi$_{2}$Te$_{2}$Se relative to \BiTe that is the source of the difficulty, suggesting that even closely related materials are not necessarily equivalent from the standpoint of thermoelectric performance.

We note that, presumably due to the weakly bonded van der Waals layers in both these materials, the lattice parameters determined from a first principles optimization can differ significantly from the
experimental lattice parameters used in the foregoing calculations (See Table 1 for the actual values). Given this, it is natural to perform an assessment of the effects of such differences on electronic structure and on
the transport quantities depicted in the above plots.  We depict such a comparison in Fig. 9 above, for the planar thermopower at 300 K.
\begin{figure}
\includegraphics[height=0.9\columnwidth,angle=270]{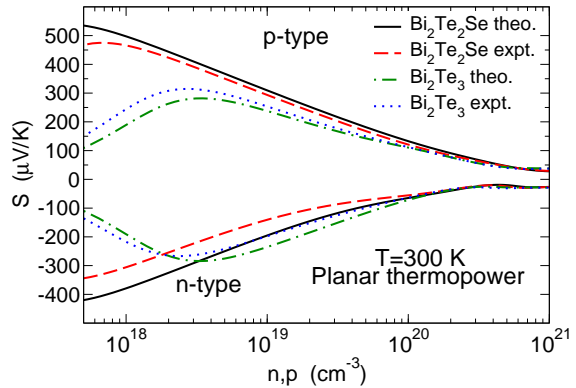}
\caption{Planar thermopower values for Bi$_2$Te$_3$ and Bi$_2$Te$_2$Se, using both the experimental and theoretical lattice parameters.}
\end{figure}
For $p$-type, the results depict a marginal decrease in \BiTe thermopower, and equally marginal increase in Bi$_{2}$Te$_{2}$ thermopower; the main effect of the smaller theoretical lattice parameters is in fact an increase in the calculated band gap of Bi$_{2}$Te$_{2}$Se by approximately 0.06 eV.  This change, however, only affects the thermopower for Bi$_{2}$Te$_{2}$Se at dopings around 10$^{18}$cm$^{-3}$, far below optimal doping, so for the purposes of assessing thermoelectric performance the effects on $p$-type of the theoretical lattice parameters are essentially nil.  With regards to $n$-type, the effects of the experimental lattice parameters are somewhat larger, but are of similar magnitude (and the same sign) for both materials, so that on a comparative basis here too the effects are rather small.  Finally, we note use of the experimental lattice parameters generally gives better agreement with experiment in these van der Waals materials and so retain their usage for the electronic structure calculations presented here.

\section{Lattice dynamics calculations}

Lattice dynamics, or phonon band structure and transport, ultimately determines the lattice thermal conductivity, a key quantity affecting thermoelectric performance.  To this end we have performed lattice dynamics calculations for Bi$_{2}$Te$_{2}$Se, using density
functional theory in Bl$\ddot{\rm{o}}$chl's
projector augmented-wave (PAW) method within the LDA as implemented in VASP. A
3$\times$3$\times$3 \emph{k}-point grid in a 3$\times$3$\times$3
supercell was used, along with an energy cutoff of 300 eV. Cell
parameters and internal coordinates were both relaxed until internal
forces were less than 2 meV/\AA. The optimized lattice constants for Bi$_{2}$Te$_{2}$Se are
 a=4.265 \AA\ and c=29.328 \AA.

In Figures 10 and 11 we present the phonon band structure and site-projected density of states for Bi$_{2}$Te$_{2}$Se.  Note that in Figures 10 and 11 we also include a band structure and density-of-states for \BiTe calculated from one of our previous works, using the same methods.  We immediately note a great similarity in the phonon bandstructures, with the main difference being slightly larger frequencies in \BiTeSe and a somewhat larger gap in the 2 - 2.5 Thz region in Bi$_{2}$Te$_{2}$Se.  It is noteworthy that the phonon band structures are so similar while the electronic band structures are so different.  Part of this is that phononic transport \begin{figure}
\includegraphics[height=0.55\columnwidth,]{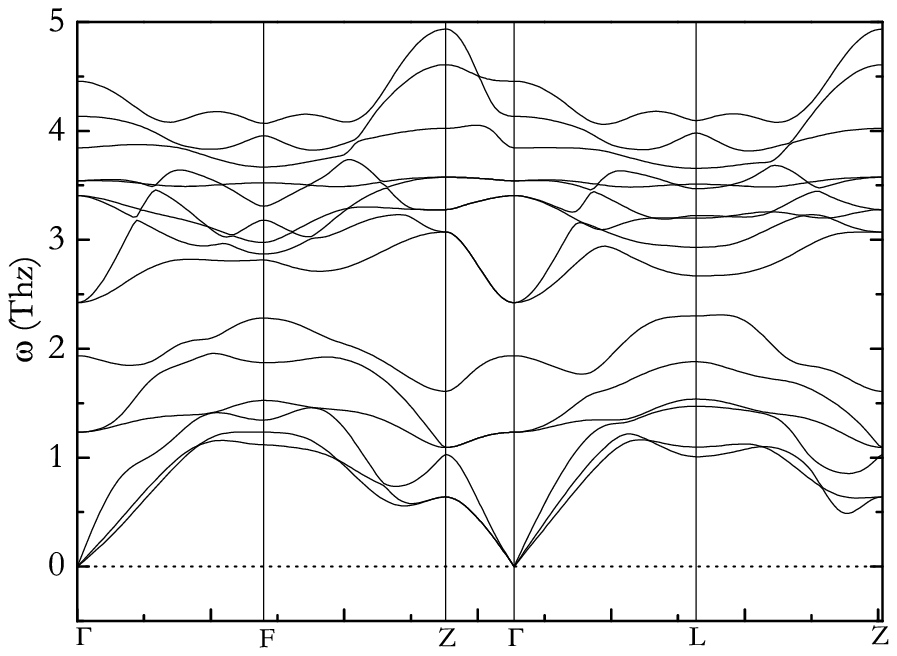}
\includegraphics[height=1.034\columnwidth,angle=270]{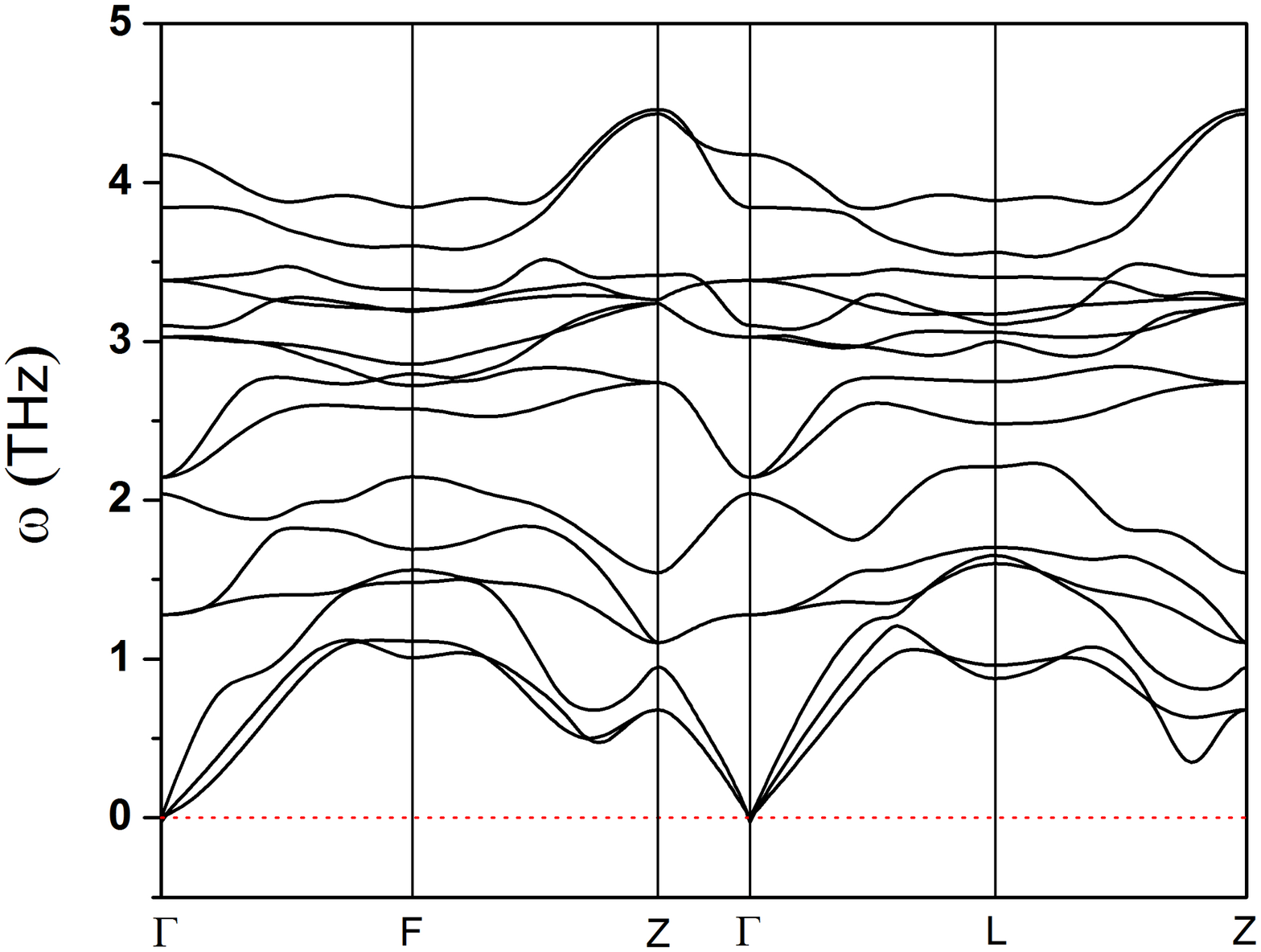}
\caption{Computed phonon band structure for \BiTeSe (top) and \BiTe (bottom) from Ref. \onlinecite{chen-bi2te3}.Coordinate of the high symmetry points are (in
units of the rhombohedral lattice vector) L:(1/2,0,0); F(1/2,0,1/2);
Z:(1/2,1/2,1/2). }
\label{pho}
\end{figure} tends to be less variable than electronic, but a more fundamental reason is that for thermoelectrics and topologically insulators, only the region near the band extrema is of relevance and these can clearly vary more widely than the entire electronic structure.  The sound speeds for \BiTeSe are somewhat higher than for \BiTe - in the nearly planar $\Gamma$-L direction the \BiTeSe sound speeds (transverse modes first) are 1524, 1763 and 2500 m/sec while the corresponding values for \BiTe are 1395, 1728, and 2394 m/sec.  For the c axis $\Gamma-$Z direction the values for \BiTeSe are 1781 (degenerate transverse mode) and 1994 m/sec, and the corresponding values for \BiTe are 1774 and 1811 m/sec.  The significantly lighter mass of Se relative to Te is likely responsible for the higher phonon frequencies and sound speeds of Bi$_{2}$Te$_{2}$Se.  

Given the higher sound speeds, the lattice thermal conductivity of \BiTeSe may be somewhat higher than that of Bi$_{2}$Te$_{3}$.  
\begin{figure}
\includegraphics[height=0.6\columnwidth]{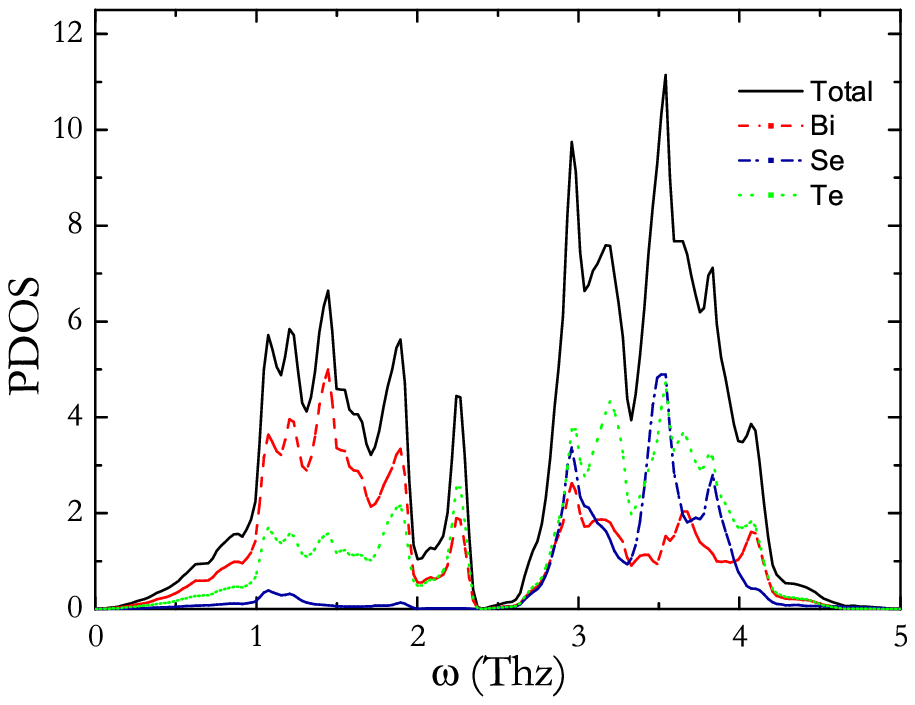}
\includegraphics[width=0.8\columnwidth,angle=270]{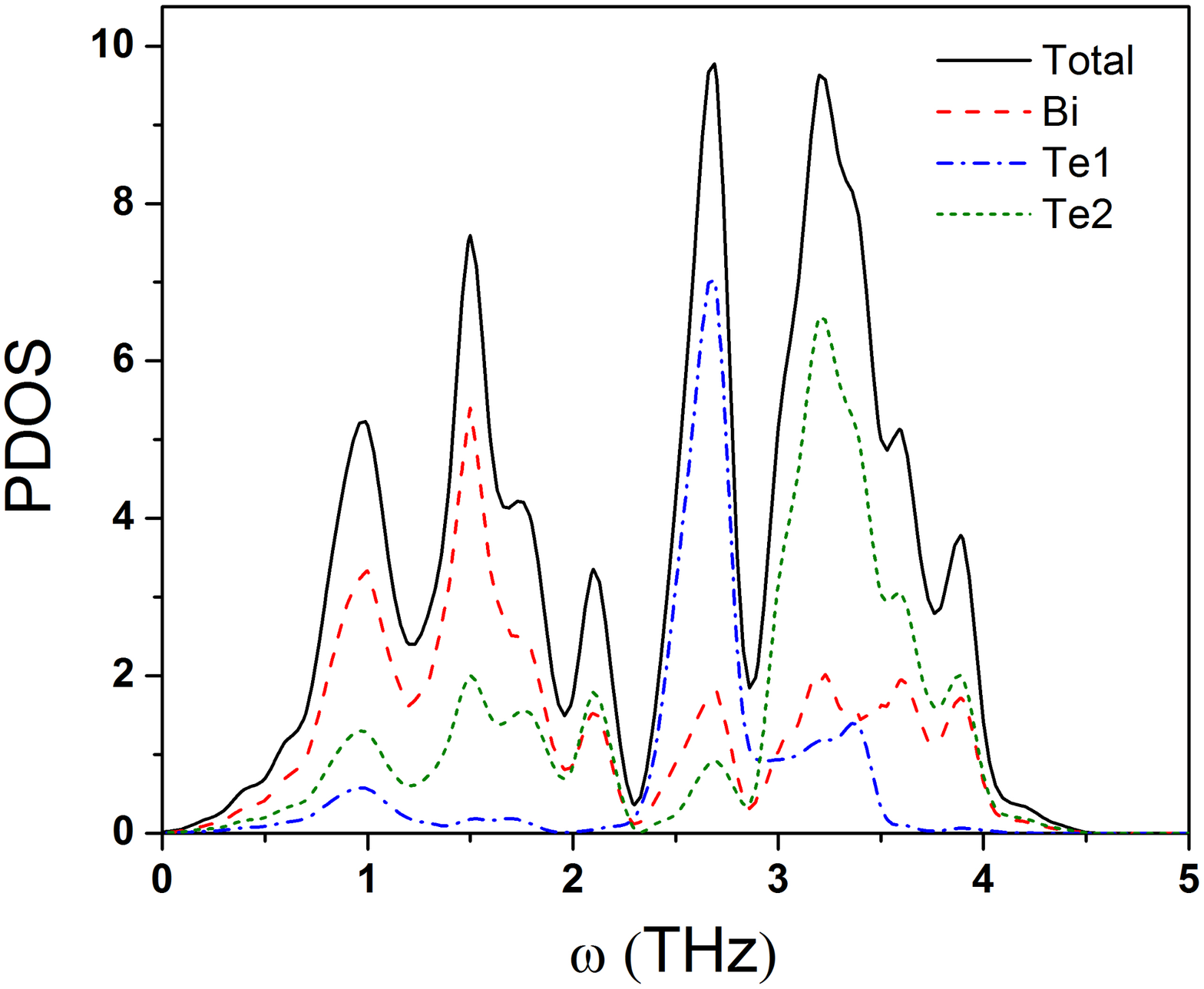}
\caption{Computed phonon density-of-states for \BiTeSe (top) and Bi$_{2}$Te$_{3}$.}
\label{pho}
\end{figure}
Note, however, that the sound speeds for \BiTeSe are still comparatively low, so that fairly low lattice thermal conductivity can be expected in the \BiTeSe material.  The lower longitudinal c-axis sound speed would suggest somewhat lower thermal conductivity in this direction than in plane.

Moving to the calculated phonon density-of-states (Fig. 11), we note immediately the prominent Se peak around 3.5 Thz, near the top end of the spectrum.  This is reasonable considering the lower mass of Se relative to Te and Bi.  The lower frequency modes below 2.5 Thz are most predominantly Bi, which again comports with the extremely heavy mass of Bi.  One final point of interest is that there is a nearly complete gap opened around 2.5 Thz.  This gap is more prominent than in \BiTe and this is again likely a result of the lighter Se atom increasing the frequency of the highest modes found between 2.5 and 4 Thz.  This also can be seen in Figure 10, where for \BiTeSe there is a gap of approximately 0.3 Thz at the $\Gamma$ point but essentially no gap at this point in Bi$_{2}$Te$_{3}$.

It is of interest to compare the behavior of the Se atom partial DOS in Bi$_{2}$Te$_{2}$Se in Fig. 11 (top) with that of Te1 in Bi$_{2}$Te$_{3}$ in Fig. 11 (bottom), since these two atoms occupy the same between-layer site (see Fig. 2). As the Figure indicates, the Te1 DOS is almost entirely (excepting the acoustic regime) comprised of a single peak around 2.7 GHz, while the Se DOS is comprised of three separate peaks at 3, 3.5 and 3.8 GHz.  All these Se peaks energies are higher than the Te1 peak in Bi$_{2}$Te$_{3}$,, as expected given the lighter mass of Se, but the split in these Se peaks is of interest.  We suspect its origin is the effectively more complex physical structure of Bi$_{2}$Te$_{2}$Se in containing three distinct atomic masses rather than two, which splits what would otherwise be a more singular peak.  

\section{Defect energy calculations and phase stability}

It is well known that \BiTe tends to form off stoichiometry due to low anti site defect formation energies.  Within this context it is of interest to consider
the defect formation energies in \BiTeSe as these will provide important information about the nature and magnitudes of defect formation, and associated scattering, in this material.  We limit
ourselves to Se/Te antisite defects as due to the equivalent charge count these energies are expected to be especially low. 

These defect calculations, as with the lattice dynamics calculations,  are based upon density
functional theory in the framework of Bl$\ddot{\rm{o}}$chl's
projector augmented-wave (PAW) method within the local density
approximation (LDA) as implemented in VASP.  We use a 4$\times$4$\times$1 conventional hexagonal
unit cell containing 240 atoms, and the 2$\times$2$\times$1
Monkhorst-Pack \emph{k}-point grid together with an energy cutoff of
500 eV.  The force convergence criterion acting on atoms is less than
0.01 eV/\AA.  The experimental lattice constants are used for
Bi$_{2}$Te$_{2}$Se, Bi$_{2}$Te$_{3}$, Bi$_{2}$Se$_{3}$, Bi, Se, and
Te as listed in Table \ref{tab:table1}.

\begin{table}
\caption{\label{tab:table1}The lattice constants we use
in this work.}
\begin{ruledtabular}
\begin{tabular}{cccccccccc}
          &Bi$^{a}$ &Se$^{b}$& Te$^{c}$&Bi$_{2}$Se$_{3}$$^{d}$  & Bi$_{2}$Te$_{3}$ & Bi$_{2}$Te$_{2}$Se\\
\hline \\
\emph{a} Experimental & 4.546   & 4.368 &4.458 &   4.135             & 4.379            & 4.305\\
\emph{c}  Experimental & 11.862  & 4.958 &5.925 &  28.615             & 30.481           & 30.00\\
\emph{a} Theoretical & -----  & ----- &---- &   ----   & 4.350  & 4.265\\
\emph{c}  Theoretical & ----  & -----  & ----- & ---- & 29.82 & 29.33 \\

\end{tabular}
$^a$experimental value in Ref.\cite{r1}\\
$^b$experimental value in Ref.\cite{r2}\\
$^c$experimental value in Ref.\cite{r3}\\
$^d$experimental value in Ref.\cite{r4}
\end{ruledtabular}
\end{table}

For the defect calculations, the formation energies $\Delta H$ for defect in
the charge state \emph{q} are given by

\begin{eqnarray}
\Delta H_{D,q}(E_{F},\mu)=&(E_{D,q}-E_{H})+\sum_{\alpha}
n_{\alpha}(\Delta \mu_{\alpha}+\mu_{\alpha}^{solid})\nonumber\\
&+q(E_{v}+E_{F}).
\end{eqnarray}
Since we only concern ourselves with the Se$_{\rm{Te}}$ and Te$_{\rm{Se}}$ antisite
defects with the same valence states, $q$ equals 0. In the first
term, $E_{D,q}$ and $E_{H}$ are the total energies of a solid with and
without defect D, respectively. The second term represents the
energy of the atom of species $\alpha$ added ($n_{\alpha}$=-1) or
removed ($n_{\alpha}$=1) from a reservoir of that species with
chemical potential $\mu_{\alpha} = \Delta
\mu_{\alpha}+\mu_{\alpha}^{solid}$.

Under equilibrium conditions for the crystal growth, the chemical
potentials $\mu_{\alpha}$ must satisfy certain conditions in order to
form a stable host compound. Other competing phases (including
elemental solids) must be avoided. In order to maintain the
stability of Bi$_{2}$Te$_{2}$Se during growth and avoid competing
phases (e.g., Bi, Te, Se, Bi$_{2}$Te$_{3}$ and Bi$_{2}$Se$_{3}$),
the relative chemical potential $\Delta \mu_{\alpha}$ must
satisfy the following limits.

\begin{eqnarray}
&2\Delta \mu_{\rm{Bi}} + 2\Delta \mu_{\rm{Te}} + \Delta
\mu_{\rm{Se}} = \Delta {\rm{H(Bi_{2}Te_{2}Se}})=- 1.478 {\rm eV},\\
&\Delta \mu_{\rm{Bi}}\leq 0, \Delta \mu_{\rm{Te}}\leq 0, \Delta \mu_{\rm{Se}}\leq 0,\\
&2\Delta \mu_{\rm{Bi}} + 3\Delta \mu_{\rm{Te}}\leq \Delta
H(\rm{Bi_{2}Te_{3}})=-1.123 eV, \\
&2\Delta \mu_{\rm{Bi}} + 3\Delta \mu_{\rm{Se}}\leq \Delta
H(\rm{Bi_{2}Se_{3}})=-1.964 eV.
\end{eqnarray}
All calculated heats of formation of ternary and binary compounds
in this work are given for per formula unit.

Eqs. 5-8 are projected to the two dimensional panel with two independent
variables, $\Delta \mu_{\rm{Te}}$ and $\Delta \mu_{\rm{se}}$ as
shown in Figure 12. The thermodynamically stable ranges of
chemical potentials of the elements in Bi$_{2}$Te$_{2}$Se
(trapezium, \emph{ABCD}) are obtained by excluding the regions of
chemical potentials in which competing phases are thermodynamically
stable as shown in Figure 12.

\begin{figure}
\includegraphics*[height=6.8cm,keepaspectratio]{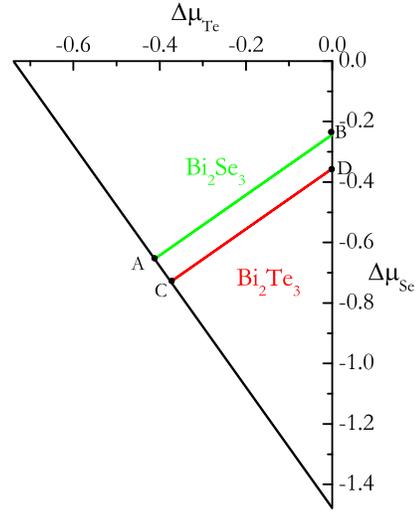}
\caption{\label{fig1}Calculated ranges of chemical potentials of the
elements involved in Bi$_{2}$Te$_{2}$Se and related competing
phases. The range of thermodynamical stability of Bi$_{2}$Te$_{2}$Se
is defined by the trapezoid ABCD.}
\end{figure}

Our calculated formation energies of antisite Se$_{\rm{Te}}$ and
Te$_{\rm{Se}}$ are are collected in Table 2, with
relative chemical potentials at the corresponding \emph{A}, \emph{B},
\emph{C}, and \emph{D} points in Figure 12.

\begin{table}
\caption{\label{tab:table2}The calculated defect formation energies
for antisite defects Se$_{\rm{Te}}$ and Te$_{\rm{Se}}$ with chemical
potentials at \emph{A}, \emph{B}, \emph{C}, \emph{D} points.}
\begin{ruledtabular}
\begin{tabular}{cccccccccc}
    &  A & B & C & D  \\
\hline \\
($\Delta \mu_{\rm{Te}}$,$\Delta \mu_{\rm{Se}}$) &(-0.41,-0.65) &(0, -0.24)&(-0.37,-0.73)&(0, -0.36) \\
Se$_{\rm{Te}}$& 0.115& 0.115 & 0.228 & 0.228  \\
Te$_{\rm{Se}}$& 0.154 & 0.154 & 0.041&  0.041   \\
\end{tabular}
\end{ruledtabular}
\end{table}

Figure 12 asserts that \BiTeSe is only thermodynamically stable within
a narrow Te-Se compositional range, above which Bi$_2$Se$_3$ would 
be formed and below which \BiTe would be performed. From Table II, we see
that certain defect structures, such as Te$_{Se}$ have defect energies as low as 0.041 eV.  When one considers
putative synthesis conditions of 1000 K, this would in equilibrium yield a Te$_{Se}$ defect concentration of order 50 percent, an absurdly large number. Hence it will be important to 
synthesize under conditions towards the Se-rich side.  Even here, though, the defect formation energies are low - 
0.115 eV for Se$_{Te}$ defects and 0.154 eV for Te$_{Se}$ defects, both less than twice the thermal energy at 1000 K, so that substantial
numbers of defects are likely to be formed at typical synthesis conditions.

There are two main points to be gleaned from these results.  Firstly, since the defect formation energies are small and asymmetric, substantial numbers of defects will form and the number of Te$_{Se}$ and Se$_{Te}$ defects will not be equal, so the material will likely form off stoichiometry.    Secondly, and more importantly, since the electronic structures of the two compounds \BiTe and \BiTeSe are so different, these large numbers of defects are likely to induce substantial alloy scattering, which is likely to significantly impair mobility.  It may also reduce the lattice thermal conductivity, but given that this is already likely to be fairly low, the mobility reduction is likely to be the larger effect.  

\section{Conclusion}

Topological insulators, such as \BiTe and \BiTeSe considered in this work, of necessity have complex band structures
due to the band inversion central to the topologically insulating behavior.  These complex band structures, in particular
highly non-parabolic isoenergy surfaces, are also those favored by high performance thermoelectrics, and these two studied materials appear to 
contain such anisotropic features, though rather different in the specifics.  The relationship between thermoelectric performance and TI behavior is
thus through the band structure as it relates to transport.  TI materials necessarily have highly non-parabolic shapes that generally
lead to corrugated isoenergy surfaces at the doping levels of interest for thermoelectrics.  These corrugated surfaces are favorable
for obtaining the combination of high conductivity and high thermopower required in a high performance thermoelectric.

The favorability of complex non-parabolic band structures for {\it both} TI behavior and high thermoelectric performance
suggests that future searches for such technologically promising materials may benefit from a consideration of the degree of complexity and anisotropy of the electronic structure
of materials studied.    It will be of interest to pursue these potentially useful behaviors from this perspective.
\\
\\
{\bf Acknowledgment}
\\
This work was sponsored by the U.S. Department of Energy (DOE) , Basic Energy Sciences, Materials Science and Engineering Division (HS and MHD), and the
DOE S3TEC Energy Frontier Research Center (DP and DJS).

\end{document}